\documentstyle[eqsecnum,aps,epsfig]{revtex}

\begin{document}
\twocolumn[\hsize\textwidth\columnwidth\hsize\csname @twocolumnfalse\endcsname
\title{Comment on "Quantum Monte Carlo Evidence for Superconductivity
in the Three-Band Hubbard Model in Two Dimensions" }

\author{H. Endres$^{1}$, W. Hanke$^{1,2}$, H.G. Evertz$^{1}$
        and F.F. Assaad$^{2}$ \\
      $^1$Physikalisches Institut, Universit\"at W\"urzburg
           97074 W\"urzburg, FRG. \\
      $^2$Department of Physics, University of California 
           Santa Barbara, CA 93106-9530 }
\maketitle
]

\narrowtext
In a recent Letter, Kuroki and Aoki \cite{Kuroki}  presented quantum
Monte-Carlo (QMC) results for pairing correlations  in the three-band
Hubbard model,  which describes the  $Cu$-$d_{x^2 - y^2}$ and $O$-$p_{x,y} $ 
orbitals present in the $CuO_2$ planes of high-$T_c$ materials.  In this
comment, we concentrate on the parameter set: 
$U_d = 3.2 t_{pd}, \Delta = 2.7 t_{pd}, t_{pp} = -0.4 t_{pd}$.   For
this parameter choice, Kuroki and Aoki see
a maximal increase in the $d_{x^2  - y ^2}$ pairing correlations which
they associate with a signature of off-diagonal long-range order
(ODLRO). We argue that:

i) The above parameter set is not appropriate for the description of high-
$T_c$ materials since it does {\it not satisfy the minimal requirement of  
a charge-transfer gap at half-filling}. To illustrate this 
point, we have calculated with QMC methods the average hole 
number as a function
of the chemical potential: $\langle n \rangle (\mu)$. Our 
results, which are plotted in Fig.1., show 
a vanishingly small charge-transfer gap (i.e. $\Delta_{ct} < 0.07
t_{pd}$). In contrast, for a physical parameter set \cite{Dopf},
one obtains a sizeable charge-transfer gap which is detectable from the
plateau in the $\langle n \rangle (\mu)$ curve (see inset in Fig. 1).
For the latter parameter set, a number of  normal state properties 
were shown to  successfully reproduce experimental data \cite{Dopf1}. 
However, despite intensive numerical efforts, no ODLRO was unambiguously
detected \cite{Assaad}.

ii) The {\it observed increase in the $d_{x^2  - y^2}$ channel }
(Fig. 2 in ref \cite{Kuroki}) {\it is dominantly 
produced by the pair-field correlations without the vertex part}
\cite{Scalapino}.
To prove this point, we have calculated 
the pair-field correlations in the d-wave channel summed over
distances $\bf{r}$ with $ |r_x|, |r_y| < R $ ($ S_d(R)$)
(See Fig. 2a). As in ref
\cite{Kuroki}, an  increase as a function of $R$ can be seen. 
However, the vertex contribution  to the
pair-field correlations, which is the relevant quantity,
is an order of magnitude smaller and shows 
- within our numerical accuray - no significant increase as a function of
$R$ (see Fig. 2b).
Hence, the claim of evidence of ODLRO is not justified.

\subsubsection*{Figure captions}
\newcounter{bean}
\vspace*{-0.3cm}\epsfig{file=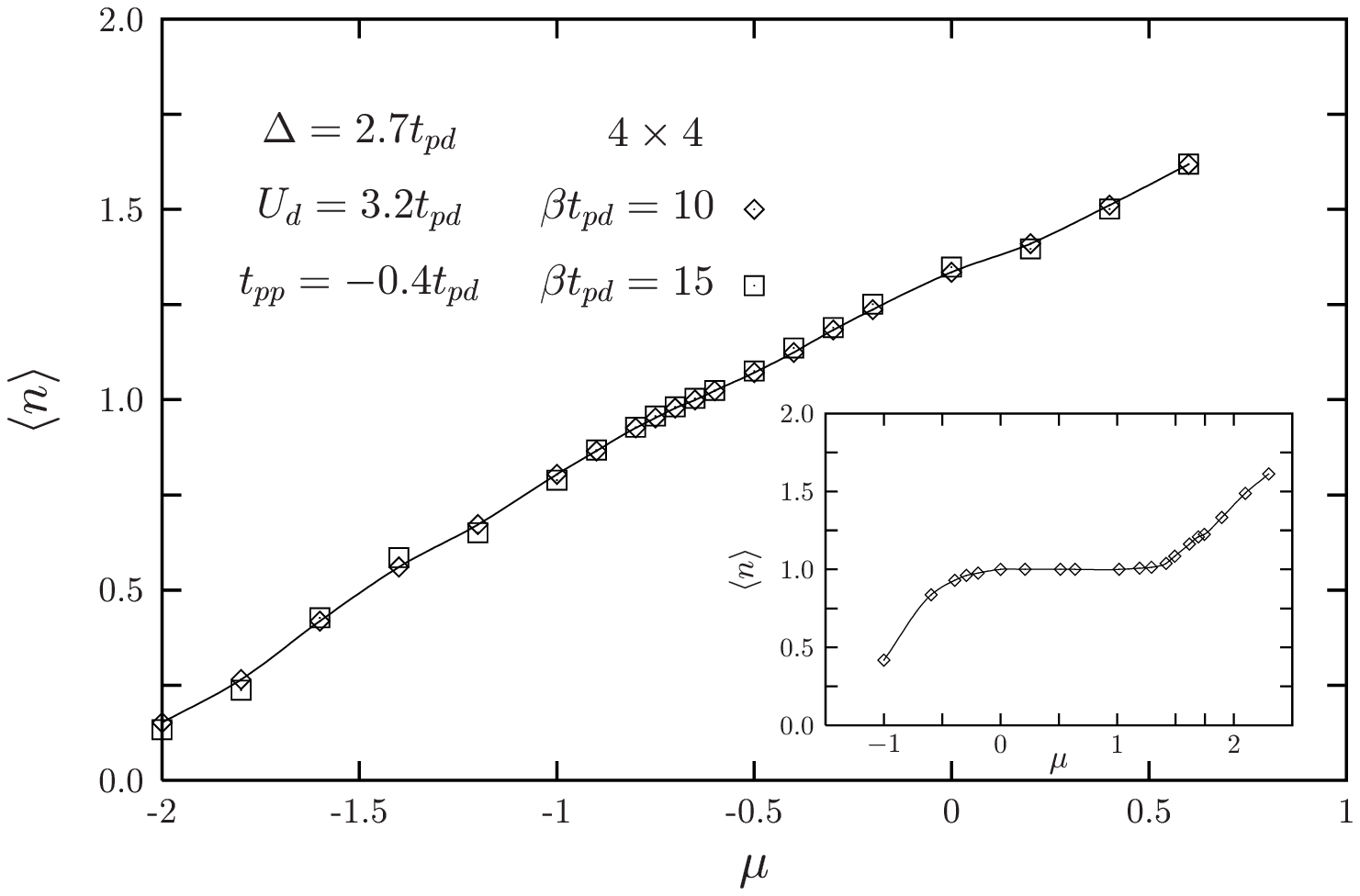, width=7.6cm}
\begin{list}%
{Fig. \arabic{bean}}{\usecounter{bean}
                   \setlength{\rightmargin}{\leftmargin}}
\item[Fig.\ 1]  $\langle n \rangle (\mu)$ on a $4\times 4$ lattice at 
$\beta t_{pd} = 10,15$ for the parameter set: 
$ U_d = 3.2 t_{pd}, \Delta = 2.7 t_{pd}, t_{pp} = -0.4 t_{pd}$. \\
Inset: $ \langle n \rangle (\mu)$ on a $4\times 4$ lattice at
$\beta t_{pd} = 10$ for $ U_d = 6 t_{pd}, \Delta = 4 t_{pd}, t_{pp} = 0.0$.
\cite{Dopf}
\vspace*{0.3cm}
\end{list}
\centerline{\epsfig{file=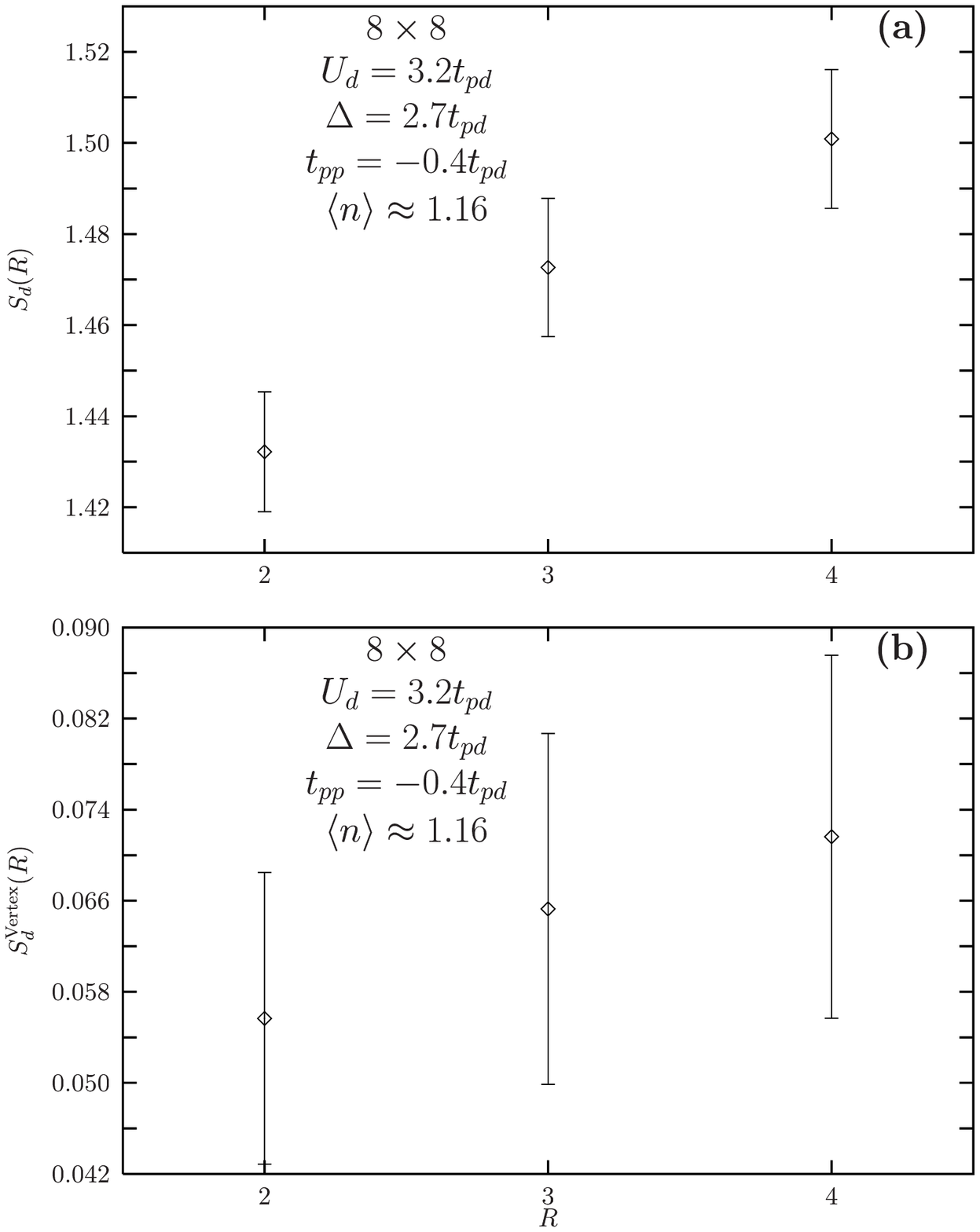, width=6.8cm}}
\begin{list}%
{Fig. \arabic{bean}}{\usecounter{bean}
                   \setlength{\rightmargin}{\leftmargin}}
\item[Fig.\ 2] 
a) $S_d(R) $ on an $ 8 \times 8$ lattice at zero temperature and
$ U_d = 3.2 t_{pd}, \Delta = 2.7 t_{pd}, t_{pp} = -0.4 t_{pd}$.
b) Vertex contribution  to $S_d (R)$ shown in a)
\end{list}


\begin{thebibliography}{99}
\bibitem{Kuroki} K. Kuroki and H. Aoki, Phys. Rev Lett. {\bf 76}, 440
(1996).
\bibitem{Dopf}  G. Dopf, A. Muramatsu and W. Hanke, Phys. Rev. Lett.
{\bf 68}, 353, (1992).
\bibitem{Dopf1} G.Dopf, J. Wagner, P. Dieterich, A. Muramatsu and W.
Hanke, Phys. Rev. Lett. {\bf 68}, 2082, (1992).
\bibitem{Assaad} F.F. Assaad, W. Hanke and D.J. Scalapino,  Phys. Rev. B
{\bf 49}, 4327 (1994).
\bibitem{Scalapino} S.R. White, D.J. Scalapino, R.L. Sugar, N.E.
Bickers and R.T. Scalettar,  Phys. Rev. B{\bf 39}, 839, (1989).
\end{thebibliography}
\end{document}